\documentclass[twocolumn,aps,prl,superscriptaddress,longbibliography]{revtex4-2}
\usepackage[colorlinks=true, citecolor=blue, urlcolor=blue, linkcolor=red]{hyperref}
\renewcommand{\section}[1]{{\par\it #1.~~}\ignorespaces}
\usepackage{amsmath,amssymb,tikz,scalerel}
\usepackage{cases}
\usepackage{bm}
\usetikzlibrary{svg.path}
\definecolor{orcidlogocol}{HTML}{A6CE39}
\tikzset{orcidlogo/.pic={
		\fill[orcidlogocol] svg{M256,128c0,70.7-57.3,128-128,128C57.3,256,0,198.7,0,128C0,57.3,57.3,0,128,0C198.7,0,256,57.3,256,128z};
		\fill[white] svg{M86.3,186.2H70.9V79.1h15.4v48.4V186.2z}
		svg{M108.9,79.1h41.6c39.6,0,57,28.3,57,53.6c0,27.5-21.5,53.6-56.8,53.6h-41.8V79.1z M124.3,172.4h24.5c34.9,0,42.9-26.5,42.9-39.7c0-21.5-13.7-39.7-43.7-39.7h-23.7V172.4z}
		svg{M88.7,56.8c0,5.5-4.5,10.1-10.1,10.1c-5.6,0-10.1-4.6-10.1-10.1c0-5.6,4.5-10.1,10.1-10.1C84.2,46.7,88.7,51.3,88.7,56.8z};}}
\newcommand\orcid[1]{\href{https://orcid.org/#1}{\mbox{\scalerel*{\begin{tikzpicture}[yscale=-1,transform shape]\pic{orcidlogo};\end{tikzpicture}}{|}}}}

\begin{document}
\title{Non-Hermitian second-order topological insulator with point gap}
\author{Xue-Min Yang\orcid{0000-0002-6937-8402}}
\affiliation{School of Electronic Science and Engineering, Chongqing University of Posts and Telecommunications, Chongqing 400065, China}
\affiliation{Chongqing Key Laboratory of Dedicated Quantum Computing and Quantum Artificial Intelligence, Chongqing 400065, China}
\affiliation{Key Laboratory of Quantum Artificial Intelligence and New Materials,Chongqing}
\affiliation{Institute for Advanced Sciences, Chongqing University of Posts and Telecommunications, Chongqing 400065, China}
\author{Hao Lin}
\affiliation{School Of Cyber Security and Information Law, Chongqing University of Posts and Telecommunications, Chongqing 400065, China}
\author{Jian Li}
\affiliation{School of Electronic Science and Engineering, Chongqing University of Posts and Telecommunications, Chongqing 400065, China}
\affiliation{Chongqing Key Laboratory of Dedicated Quantum Computing and Quantum Artificial Intelligence, Chongqing 400065, China}
\affiliation{Key Laboratory of Quantum Artificial Intelligence and New Materials,Chongqing}
\affiliation{Institute for Advanced Sciences, Chongqing University of Posts and Telecommunications, Chongqing 400065, China}
\author{Jia-Ji Zhu}
\affiliation{School of Electronic Science and Engineering, Chongqing University of Posts and Telecommunications, Chongqing 400065, China}
\affiliation{Chongqing Key Laboratory of Dedicated Quantum Computing and Quantum Artificial Intelligence, Chongqing 400065, China}
\affiliation{Key Laboratory of Quantum Artificial Intelligence and New Materials,Chongqing}
\affiliation{Institute for Advanced Sciences, Chongqing University of Posts and Telecommunications, Chongqing 400065, China}
\author{Jun-Li Zhu}
\affiliation{School of Mathematics and Statistics, Chongqing University of Posts and Telecommunications. }
\author{Hong Wu\orcid{0000-0003-3276-7823}}\email{Contact author: wuh@cqupt.edu.cn}
\affiliation{School of Electronic Science and Engineering, Chongqing University of Posts and Telecommunications, Chongqing 400065, China}
\affiliation{Chongqing Key Laboratory of Dedicated Quantum Computing and Quantum Artificial Intelligence, Chongqing 400065, China}
\affiliation{Key Laboratory of Quantum Artificial Intelligence and New Materials,Chongqing}
\affiliation{Institute for Advanced Sciences, Chongqing University of Posts and Telecommunications, Chongqing 400065, China}

\begin{abstract}

The zero-mode corner states in the gap of two-dimensional non-Hermitian Su-Schrieffer-Heeger model are robust to infinitesimal perturbations that preserve chiral symmetry. However, we demonstrate that this general belief is no longer valid in large-sized
systems. To reveal the higher-order topology of non-Hermitian systems, we establish a correspondence between the stable zero-mode singular states and the topologically protected corner states of energy spectrum in the thermodynamic limit. Within this framework, the number of zero-mode singular values is directly linked to the number of mid-gap corner states. The winding numbers in real space can be defined to count the number of stable zero-mode singular states. Our results formulate a bulk-boundary correspondence for both static and Floquet non-Hermitian systems, where topology arises intrinsically from the non-Hermiticity, even without symmetries.

\end{abstract}
\maketitle
\section{Introduction}
Topological phases have received extensive attention and developed rapidly due to their potential applications in functional devices\cite{RevModPhys.82.3045,RevModPhys.91.015006,RevModPhys.91.015005,RevModPhys.93.025002}. The core feature of this type of phase is the existence of protected edge states in the gap \cite{RevModPhys.82.3045}. This property can be topologically protected by the symmetries and energy gap of the system. According to the principle of bulk-boundary correspondence, topological invariants can be defined to count the number of boundary states \cite{RevModPhys.88.035005}. 

In recent years, the research of topological insulators has been extended to higher-order cases \cite{Benalcazar_2017,PhysRevLett.118.076803,PhysRevLett.121.116801,PhysRevLett.126.206404,PhysRevX.9.011012,PhysRevLett.124.036803}. The $d$ dimensional higher-order topological insulators of order $n$ possess $d-n$ dimensional boundary state, such as corner or hinge states \cite{Schindler_2018,Chen_2021,b2s4-b2sn,p91n-lvqv,zpcc-59kv}. This type of phase is protected by crystalline symmetries. They have been realized in photonic crystal slabs \cite{PhysRevLett.122.233902}, waveguide array \cite{PhysRevLett.125.213901}, electric circuits \cite{PhysRevLett.126.146802}, acoustic crystal \cite{PhysRevLett.131.157201}, and ultracold atoms \cite{dong2025observationhigherordertopologicalbound}. On the other hand, non-Hermitian topological phases have also attracted widespread attention \cite{PhysRevX.9.041015,PhysRevLett.130.266901,PhysRevB.109.205142,PhysRevLett.123.206404,PhysRevLett.130.157201,PhysRevLett.130.066601,9b46-d2ry,q3db-9zkj,2zx3-rs57,ma2025dynamically,wu2025topological,PhysRevA.100.032102,ghorashi2025topologicalrealityswitchbulkboundary}. A typical feature of such systems is the skin effect with abundant bulk states localized at the edges \cite{PhysRevLett.134.196302,PhysRevB.109.165127,q6wr-2rt9}. Therefore, edge states can no longer be characterized by the topological properties of bulk bands, indicating the breakdown of the bulk-boundary correspondence \cite{PhysRevLett.116.133903}. By replacing the Brillouin zone with the generalized Brillouin zone, topological invariants based on non-Bloch Hamiltonian can be constructed to accurately describe the number of edge states  \cite{PhysRevLett.121.086803,PhysRevLett.123.066404,PhysRevB.102.041119}. 

The combination of higher-order topology and non-Hermitian systems has led to the emergence of non-Hermitian higher-order topological insulators \cite{PhysRevLett.122.076801,PhysRevLett.122.195501,PhysRevB.102.094305}. The bulk-boundary correspondence has also been studied in these phases. However, under open boundary conditions, the energy spectrum is highly sensitive to perturbations that even maintain symmetry \cite{PhysRevLett.134.056601,wu2025breakdownnonblochbulkboundarycorrespondence,Trefethen2005}. Thus, revealing the non-Hermitian higher-order topology remains a problem. 

In this work, we systematically studied static and periodically driven second-order topological insulators. Motivated by the instability of the energy spectrum in finite-sized systems, we establish a correspondence between the number of stable zero-mode singular states and the number of protected topological corner states at the thermodynamic limit energy. The number of corner states in the thermodynamic limit is given by the new topological number instead of the non-Bloch topological number. Our results provide a promising new avenue for exploring novel non-Hermitian topological phases.

\begin{figure}[tbp]
\centering
\includegraphics[width=0.95\columnwidth]{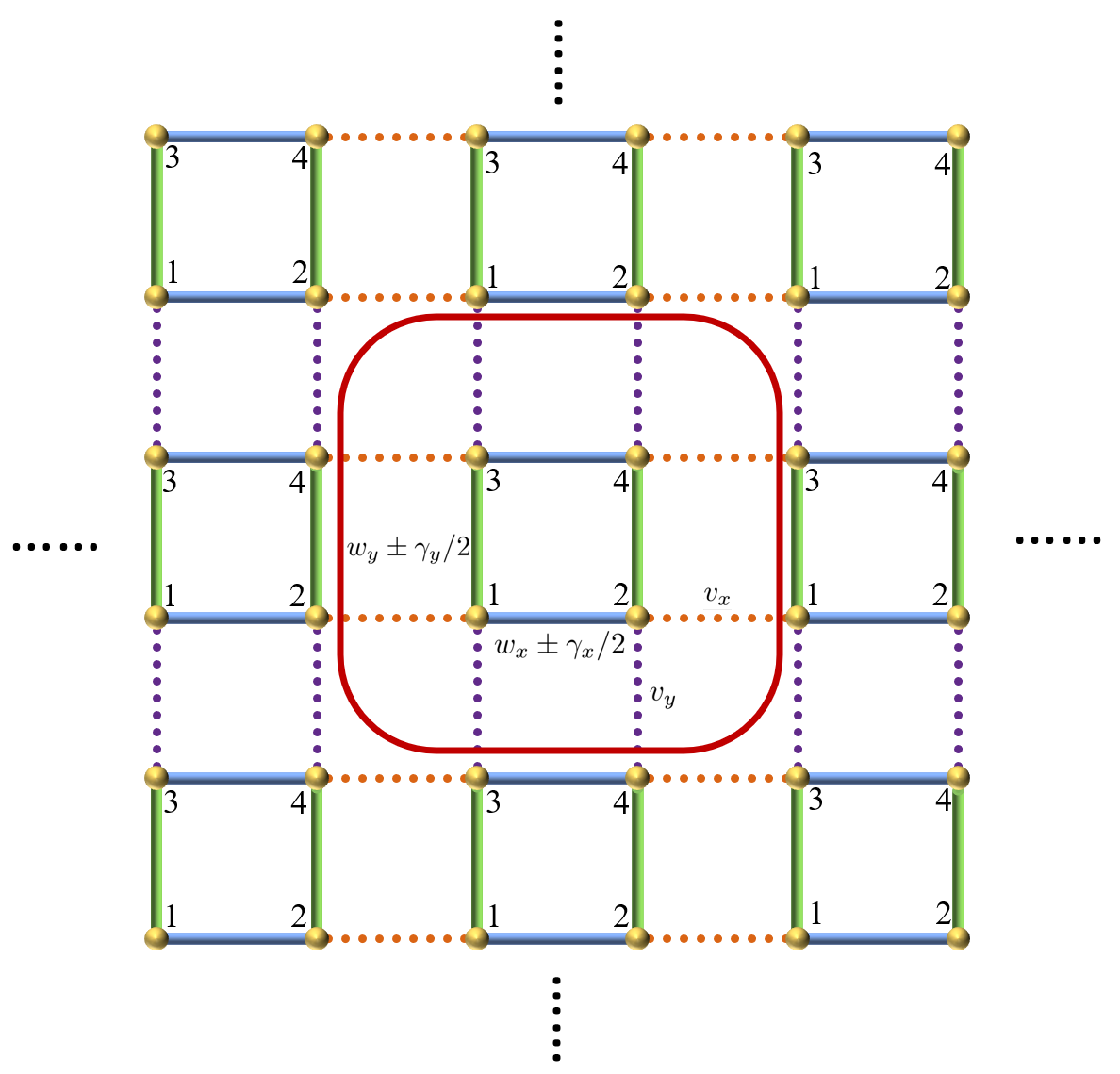}
 \caption{Schematics of the 2D non-Hermitian Su-Schrieffer-Heeger model on a square lattice.  The box indicates the unit cell.}
\label{tv11}
\end{figure}

\section{Model and Breakdown of Bulk-Boundary Correspondence}We consider a two-dimensional ($2$D) non-Hermitian Su-Schrieffer-Heeger (SSH)
model, as illustrated in Fig. %
\ref{tv11}, where both the $x$ and $y$ directions consist of $1$D non-Hermitian
SSH chains \cite{PhysRevLett.132.220402}. The non-Hermiticity arises from
asymmetric intra-cell hoppings along both directions. This model forms a superlattice with a $4$-site unit cell based
on a tensor-product structure, where each unit cell can be viewed as the
tensor product of $1$D dimers along the $x$ and $y$ directions, naturally
supporting higher-order topological corner states. The Bloch Hamiltonian reads
\begin{equation}
H_{2D}(k_{x},k_{y})=H_{x}(k_{x})\otimes \mathbb{I}_{2}+\mathbb{I}_{2}\otimes
H_{y}(k_{y}),
\end{equation}%
with each $1$D block given by
\begin{eqnarray}
H_{\alpha }(k_{\alpha }) &=&\mathbf{h}_{\alpha }(k_{\alpha })\cdot \mathbf{%
\bm{\sigma} } \\
&\mathbf{=}&(w_{\alpha }+v_{\alpha }\cos k_{\alpha })\sigma _{x}+(v_{\alpha
}\sin k_{\alpha }+i\gamma _{\alpha }/2)\sigma _{y},  \notag
\end{eqnarray}%
for $\alpha =x,y$, where $v_{\alpha }$ ($w_{\alpha }$) denotes the
inter-cell (intra-cell) hopping amplitude, and $\gamma _{\alpha }$
quantifies the degree of non-Hermitian asymmetry. $\sigma_{x,y}$ are the Pauli matrices and $\mathbb{I}_2$ is the $2\times2$ identity matrix.

The eigenvalues of this model are exactly determined as $%
E=\varepsilon _{x}(k_{x})+\varepsilon _{y}(k_{y})$, with $\varepsilon
_{\alpha }(k_{\alpha })=\pm \sqrt{(w_{\alpha }+v_{\alpha }\cos k_{\alpha
})^{2}+(v_{\alpha }\sin k_{\alpha }+i\gamma _{\alpha }/2)^{2}}$. The $2$D eigenstates $|\psi
\rangle _{2D}=|\psi _{k_{x}}\rangle \otimes |\psi _{k_{y}}\rangle $
is the tensor product of $1$D eigenstates, where $%
H_{\alpha }(k_{\alpha })|\psi _{k_{\alpha }}\rangle =\varepsilon _{\alpha
}(k_{\alpha })|\psi _{k_{\alpha }}\rangle $. The $2$D topological invariant $\mathcal{V}%
_{2D}=\mathcal{V}_{x}\mathcal{V}_{y}$ is fully determined by two independent
$1$D winding numbers:
\begin{equation}
\mathcal{V}_{\alpha }=\frac{1}{2\pi }\frac{\left[ \arg h_{\alpha
}^{+}-\arg h_{\alpha }^{-}\right] _{C_{\tilde{k}}}}{2}\in \{0,1\},
\end{equation}%
where $h_{\alpha }^{\pm }=h_{\alpha }^{x}\pm ih_{\alpha }^{y}$ and [${\arg
h_{\alpha }^{\pm }}]$$_{{C_{\tilde{k}}}}$ are the phase changes of $h_{\alpha }^{\pm }$ as $\tilde{%
k}$ counterclockwisely goes along the generalized Brillouin zone (GBZ) $C_{%
\tilde{k}}$. By  replacing the $k_x,k_y$ with $k_x-i\ln\sqrt{\lvert\frac{w_x-\gamma_x/2}{w_x+\gamma_x/2}\lvert},\,k_y-i\ln\sqrt{\lvert\frac{w_y-\gamma_y/2}{w_y+\gamma_y/2}\lvert}$, we can obtain GBZ. A higher-order topological phase with four protected corner states
emerges precisely when $\mathcal{V}_{x}=\mathcal{V}_{y}=1$. This
model is one of the standard models for investigating non-Hermitian
higher-order topology. 

\begin{figure}[tbp]
\centering
\includegraphics[width=1\columnwidth]{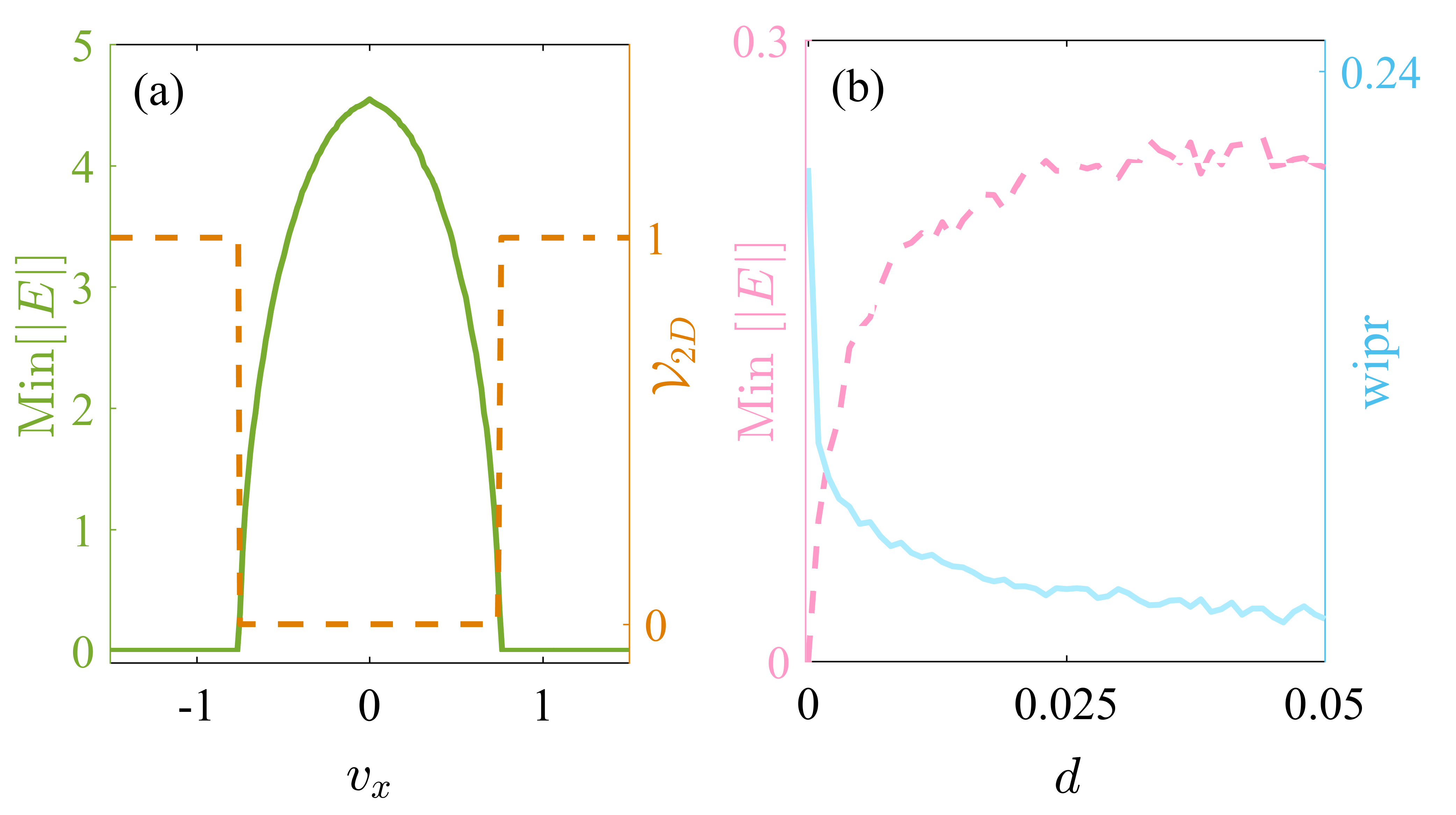}
\caption{(a) The minimum modulus of the energy eigenvalues, Min$[|E|]$ under open
boundary conditions as a function of the horizontal inter-cell hopping
amplitude $v_{x}$ (green solid curve). The orange dashed curve shows the corresponding winding number $\mathcal{V}_{2D}$. The system size is $1000\times 1000$ lattice sites, with
parameters $w_{x}=1$, $\gamma _{x}=1.5$, $w_{y}=0$, $\gamma _{y}=9$, and $%
v_{y}=6v_{x}$. When $\left\vert v_{x}\right\vert >0.75$, the system enters a
higher-order topological phase, signaled by Min$[|E|]\rightarrow 0$, indicating the emergence of topologically protected corner
states.  (b) The minimum energy modulus (pink dashed curve) and the
weighted inverse participation ratio (wipr, sky-blue solid curve) as a function of
disorder strength $d$. The system size is $60\times 60$, with fixed
parameters $v_{x}=-1.5$, $w_{x}=1$, $\gamma _{x}=1.5$, $w_{y}=0$, $\gamma
_{y}=9$, and $v_{y}=-9$. }
\label{tv22}
\end{figure}

As demonstrated in Fig. \ref{tv22}(a), when $|v_{x}|>0.75$, the smallest $|E|$ (Min$[|E|]$) equal to $0$ under open boundaries, which is a
direct signature of topologically protected corner states. Meanwhile, the
bulk winding number $\mathcal{V}_{2D}$ jumps to $1$ at the same critical points,
confirming that the bulk topological invariant accurately predicts the
boundary phenomenon. This is a manifestation of the bulk-boundary correspondence in higher-order non-Hermitian systems.
However, the bulk-boundary correspondence in this non-Hermitian system is
highly sensitive to perturbations. As shown in Fig. \ref{tv22}(b), this correspondence can still break down even under weak perturbations that preserve chiral symmetry. We add the perturbation term
\begin{equation}
\Delta H=\Delta H_{x}\otimes \mathbb{I}+\mathbb{I}\otimes \Delta H_{y},
\end{equation}%
where $\Delta H_{\alpha }=d\left( \kappa _{ij}a_{i}^{\dagger }b_{j}+\kappa
_{ji}b_{j}^{\dagger }a_{i}\right) $, with $a_{i}$ ($b_{j}$) being the
annihilation operators on the different sublattice A (B) of the $i$th ($j$%
th) unit cell in either the $x$- or $y$-direction, and $\kappa _{ij},\,\kappa
_{ji}\in \lbrack -0.5,0.5]$ are random variables modeling disorder of
strength $d$. Here, $\Gamma \Delta H\Gamma ^{-1}=-\Delta H$, with $\Gamma $
denoting the chiral operator in real space. Even if the strength of disorder is very small, the Min$[|E|]$ can deviate significantly from zero (pink dashed curve in  Fig. \ref{tv22}(b)). To further view this phenomenon, we introduce the weighted inverse participation
ratio (wipr) rather than inverse participation
ratio in non-Hermitian systems to describe skin effect. Here, wipr is given by
\begin{equation}
\mathrm{wipr}=\sum_{n,x,y}\frac{|\psi _{n,x,y}|^{4}}{4L_{x}L_{y}}\sqrt{%
\left( x-\frac{L_{x}}{2}\right) ^{2}+\left( y-\frac{L_{y}}{2}\right) ^{2}},
\end{equation}%
where $x,\,y$ are the lattice site indexes in the $x$- and $y$-direction, and $%
L_{x}$ ($L_{y}$) are the length of chains in $x$ ($y$)-direction. It characterizes the extent to which the wavefunction is localized at the boundary. As shown in Fig. %
\ref{tv22}(b), the wipr (sky-blue solid curve) decreases sharply with
disorder strength $d$, indicating that the skin effect is highly
sensitive to perturbations. Therefore, the energy spectrum in the large-sized systems is unstable. To explain this, we introduce the condition number 
\begin{equation}
\kappa=\lvert\lvert \Sigma \lvert\lvert _2 \lvert \lvert \Sigma^{-1}\lvert\lvert_2, 
\end{equation}
where $\lvert\lvert \Sigma \lvert\lvert_2$ is the 2-norm of $\Sigma$ and $\Sigma$ is a matrix of eigenvectors of Hamiltonian \cite{Trefethen2005}. For the parameters in Fig. \ref{tv22}(b), $\kappa=2.93\times 10^{9}\gg1$ shows that the Hamiltonian is in some sense far from normal. Therefore, the eigenvalues of this non-Hermitian system are highly sensitive to perturbations \cite{Trefethen2005}. The above results demonstrate that the zero-modes
can be destroyed by sufficiently weak perturbations, even when chiral
symmetry is preserved, proving that these corner states lack topological
protection. In this regime, the non-Bloch bulk invariant fails to
reliably predict the presence or nature of boundary states, signaling a
breakdown of the bulk-boundary correspondence. 

\section{Restoring Bulk-Boundary Correspondence via Singular Value Spectrum}
\begin{figure}[tbp]
\centering
\includegraphics[width=1\columnwidth]{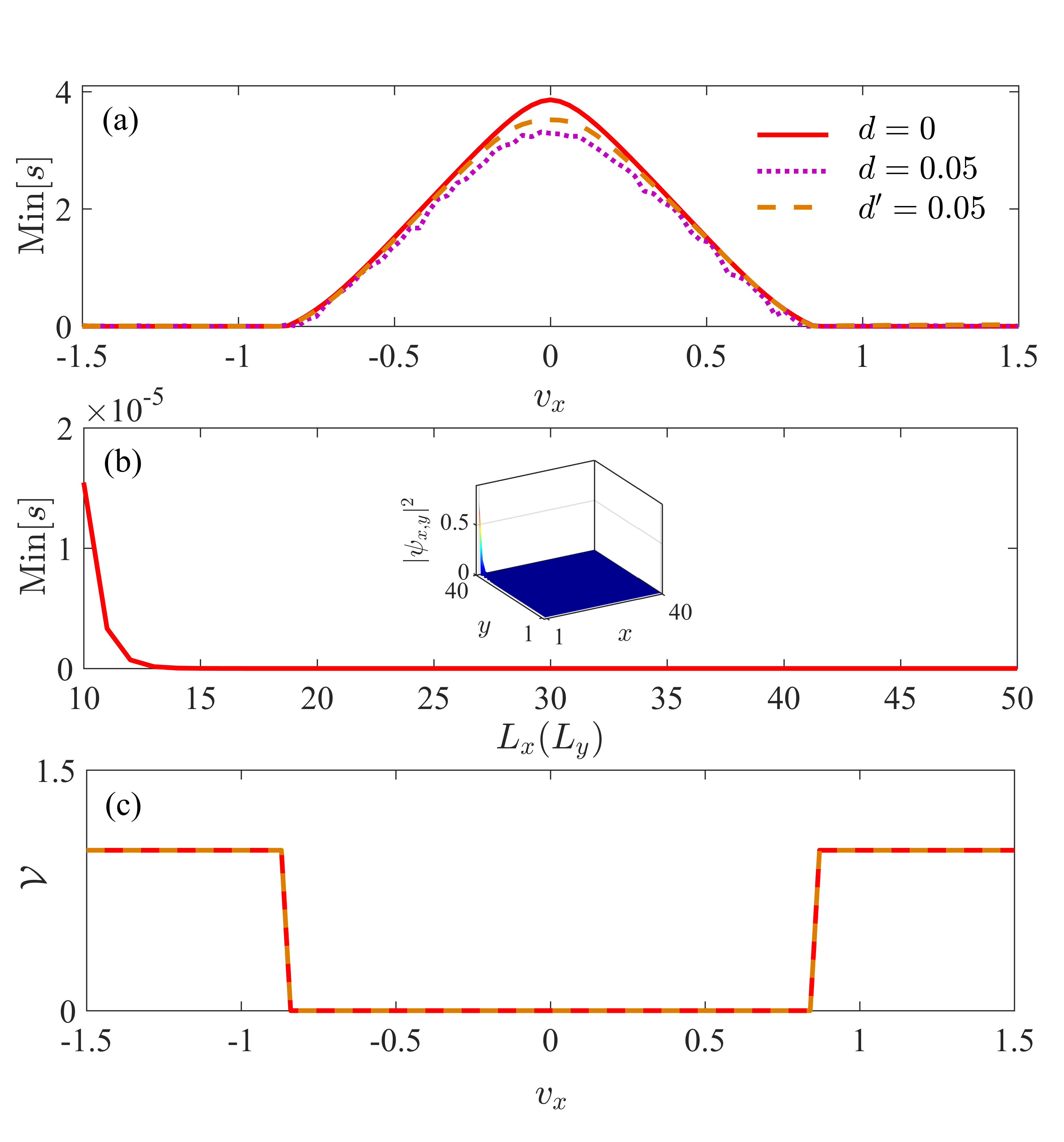}
\caption{(a) The smallest singular value (Min$[s]$) of the 2D non-Hermitian SSH model as a function of the horizontal inter-cell hopping amplitude $v_x$. The red solid curve corresponds to the clean system ($d = 0$), while the magenta dashed curve shows the result with weak disorder of strength $d = 0.05$ that preserves sublattice symmetry; the orange dashed curve represents the case where both $H_x$ and $H_y$ are perturbed by a fully random and symmetry-breaking disorder of strength $d' = 0.05$. Remarkably, the zero-mode of Min$[s]$ persists, indicating that the bulk-boundary correspondence remains intact.  (b) Exponential decay of the smallest singular value with increasing system size in either the $x$- or $y$-direction ($L_x$ or $L_y$), confirming its correspondence to a zero-mode. The inset displays the spatial probability distribution $|\psi_{x,y}|^2$ of the associated corner state, localized at the system's corner. (c) Winding number $\mathcal{V}$ of the bulk Hamiltonian as a function of $v_x$. In panels (a) and (c), the system size is $L_x=L_y=40$; all panels share the same parameters: $w_x = 1$, $w_y = 0$, $\gamma_x = 1.5$, $\gamma_y = 9$, and $v_y = 6v_x$.}
\label{tv33}
\end{figure}

Given the fact that traditional topological invariants based on the generalized Brillouin zone fail to accurately predict boundary phenomena, a method utilizing the singular value spectrum of the Hamiltonian is proposed to restore the bulk-boundary correspondence. We
establish a rigorous connection between the singular value decomposition
(SVD) of the non-Hermitian system $H_{2D}=USV^{\dagger }$ \cite{PhysRevA.99.052118} and the emergence
of zero-mode corner states in energy spectrum for themordynamic limit. Denoting the column
vectors of $V$ ($U$) by $v_{n}$ ($u_{n}$) and singular values on the
diagonal of $S$ by $s_{n}$, we can obtain
\begin{equation}
H_{2D}^{\dagger }H_{2D}v_{n}=s_{n}^{2}v_{n}.\quad
\end{equation}%
and%
\begin{equation}
H_{2D}v_{n}=s_{n}u_{n}.
\end{equation}%
when $s_{n}\rightarrow 0$ in the thermodynamic limit, the system supports
zero-mode states with wave function $v_n$. As
shown in Fig.~\ref{tv33}(a), the smallest singular value (Min$[s]$) of the
2D non-Hermitian SSH model is very small for some areas under both clean ($d=0$) and weakly disordered ($d=0.05$) conditions. In Fig.~\ref{tv33}(b), we demonstrate that such Min$[s]$ decays exponentially with increasing system size $L_{x}$ or $L_{y}$, confirming the zero-mode singular value in the thermodynamic limit. Besides, these states are all corner states (The inset of Fig.~\ref{tv33}(b) shows the probability distribution of one of the zero-mode states). This reveals that the associated eigenstate of $H_{2D}$ is also spatially localized at the system's corner. Due to the fact that the singular values of $H_{2D}$ are the square root of eigenvalues of Hermitian operator $H^{\dagger }_{2D}H_{2D}$, they are always real, non-negative, and very stable under weak perturbations \cite{Ashida_2020}. Therefore, Min$[s]$ remains insensitive to the disorder strength and size, which is a key feature enabling its use as a stable index to see corner states.  

If the non-Hermitian system $H_{2D}$ has a point gap at $E=0$, we can define a topological invariant via the
chiral-symmetric Hermitian matrix \cite{PhysRevLett.128.127601}
\begin{equation}
\tilde{H}=%
\begin{pmatrix}
0 & H_{2D} \\
H_{2D}^{\dagger } & 0%
\end{pmatrix}%
,
\end{equation}%
whose eigenvalues are $\pm s_{n}$, where $s_{n}$ represents the singular
value of $H_{2D}$. The winding number $\mathcal{V}$ of $\tilde{H}$, computed
as
\begin{equation}
\mathcal{V}=\frac{1}{4\pi i}\mathrm{Tr}\ln (P^{A}{P^{B}}^{\dag}),
\end{equation}
where $P^{A}=U^{\dag }PU$, $P^{B}=V^{\dag }PV$, and
\begin{equation}
P=\text{Diag}[e^{\frac{-i2\pi \times 1\times 1}{L_{x}L_{y}}},e^{\frac{-i2\pi
\times 1\times 2}{L_{x}L_{y}}},...e^{\frac{-i2\pi \times L_{x}L_{y}}{%
L_{x}L_{y}}}]\otimes \mathbb{I}_{2}.  \label{P}
\end{equation}
 As shown in Fig.~\ref{tv33}%
(c), the position where $\mathcal{V}$ jumps from $0$ to $1$ precisely coincides with the location where Min$[s]$ becomes zero. The total number of zero-mode singular
values equals $2\mathcal{V}$, which directly corresponds to the number of
topologically protected corner states of energy spectrum in the thermodynamic limit.

To further examine the robustness of our bulk-boundary correspondence,
we introduce a fully random and symmetry-breaking weak disorder $%
\triangle H^{\prime }=\Delta H_{x}^{\prime }\otimes \mathbb{I}+\mathbb{I}%
\otimes \Delta H_{y}^{\prime }$ to the Hamiltonian. Each matrix element of $%
\triangle H_{\alpha }^{\prime }$ is drawn independently and uniformly from
the interval $[-d^{\prime },d']$. Remarkably, despite the introduction of a fully
random, symmetry-breaking disorder, the bulk-boundary correspondence remains
intact. As shown in Fig.~\ref%
{tv33}(a), the smallest singular value Min$[s]$ ia also equal to zero in the
topologically nontrivial regime, and the number of zero-mode singular values
exactly matches the bulk winding number $2\mathcal{V}$. This demonstrates
that the singular-value-based topological characterization is robust against
perturbations that would otherwise destabilize conventional bulk-boundary correspondence. Our framework thus provides a stable and reliable route to
identify topological boundary states in non-Hermitian systems, even in the absence of chiral symmetry.

 \begin{figure}[tbp]
\centering
\includegraphics[width=1\columnwidth]{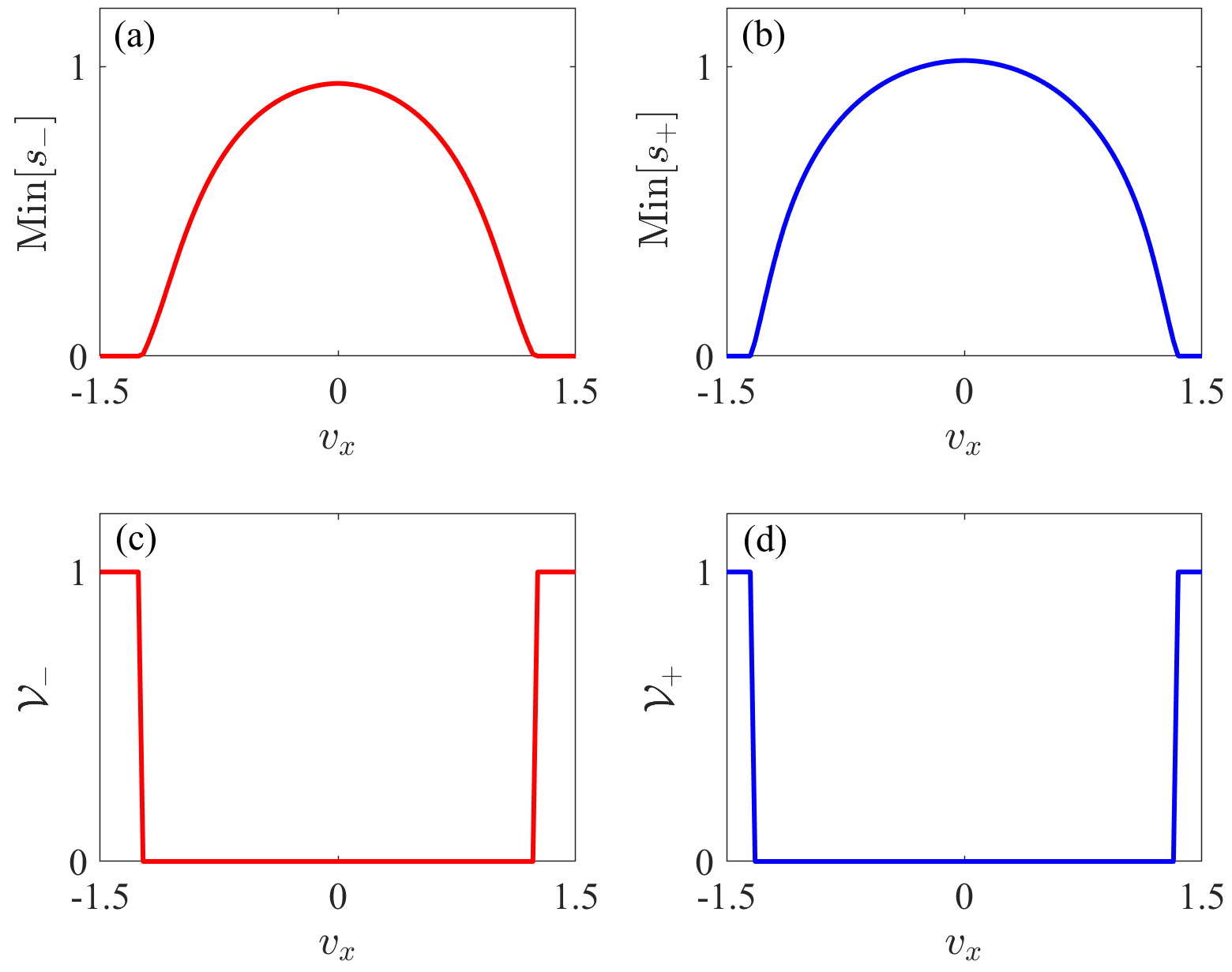}
\caption{(a-b) The smallest singular values Min$[s_{\mp}]$ of (a) $U(T) - \mathbb{I}$ and (b) $U(T) + \mathbb{I}$ as functions of the horizontal inter-cell hopping amplitude $v_x$. The zero-mode singular values in (a) and (b) signal the presence of topological $0$-mode and $\pi$-mode states, respectively. (c-d) Corresponding topological invariants $\mathcal{V}_-$ and $\mathcal{V}_+$ as functions of $v_x$, which characterize the $0$- and $\pi$-quasienergy topological phases. Parameters: system size $L_x = L_y = 40$; $w_x = 1$, $w_y = 0$, $\gamma_x = 1.5$, $\gamma_y = 10.5$, $v_y = 7v_x$; full driving period $T = 0.6$, the duration of the first segment within each period $T_1 = 0.3$, and the intensity ratio factor between the second and the first stage $q_x=q_y = 0.2$.
}
\label{tv44}
\end{figure}

\section{Extension to Floquet Non-Hermitian Systems}
The method used to establish bulk-boundary correspondence in static system can be generalized to periodically driven non-Hermitian systems. We consider a Floquet 2D Non-Hermitian SSH model with
time-periodic hopping amplitudes:
\begin{equation}
v_{\alpha}(t) =\begin{cases}f_{\alpha} ,& t\in\lbrack mT, mT+T_1)\\q_{\alpha}\,f_{\alpha},& t\in\lbrack mT+T_1, (m+1)T) \end{cases}~, \label{procotol}
\end{equation}
where $T$ is the driving period, $m\in \mathbb{Z}$, $T_{1}$ is the duration of the first
segment, and $q_{\alpha }$ controls the amplitude ratio between different segments.
This system lacks a well defined energy spectrum. According to Floquet theorem, the one-period evolution operator
$U(T)=\mathbb{T}e^{-i\int_{0}^{T}H(t)dt}$ can be used to define an effective Hamiltonian
\begin{equation}
{H}_{\text{eff}}={\frac{i}{T}}\ln [{U}(T)]=H_{\text{eff},x}\otimes
\mathbb{I}+\mathbb{I}\otimes H_{\text{eff},y},
\end{equation}
where
\begin{equation}
H_{\text{eff},\alpha }=\frac{i}{T}\ln [e^{-iH_{2}(k_{\alpha
})T_{2}}e^{-iH_{1}(k_{\alpha })T_{1}}]\,\,\,(\alpha =x,y),
\end{equation}%
and $T_{2}=T-T_{1}$. Eigenvalues of the effective Hamiltonian ${H}_{%
\text{eff}}$ lie on the $\varepsilon _{j}\in \lbrack -\pi /T,\pi
/T]$ and constitute the quasienergy spectrum of the system\cite%
{PhysRevLett.113.236803}. Topological phase of periodically driven system
are defined in such quasienergy spectrum \cite{PhysRevB.89.121401,PhysRevLett.113.236803,PhysRevB.96.155118,PhysRevLett.110.016802,Bai01012021,PhysRevB.111.195424,PhysRevB.109.035418}.

To characterize topology in this setting, we introduce two auxiliary
operators: $U(T)-\mathbb{I}$ and $U(T)+\mathbb{I}$, whose singular values
directly probe the presence of 0-mode and $\pi $-mode states,
respectively. As shown in Figs.~\ref{tv44}(a,b), the smallest singular
values Min$[s_{\mp}]$ of these operators $U(T)\mp \mathbb{I}$ equal to zero, signaling the emergence
of topologically protected $0$- and $\pi $-modes under open boundaries.
Crucially, unlike the complex quasienergy spectrum, these singular values
remain robust against disorder and insensitive to boundary conditions, which
is a key advantage for practical characterization.

We further define two topological invariants, $\mathcal{V}_{+}$
and $\mathcal{V}_{-}$, based on the winding numbers of
\begin{equation}
\tilde{H}_{\pm }^{\prime }=%
\begin{pmatrix}
0 & U(T)\pm \mathbb{I} \\
U(T)^{\dag }\pm \mathbb{I} & 0%
\end{pmatrix}%
,
\end{equation}
which are computed as
\begin{equation}
\mathcal{V}_{\pm }=\frac{1}{4\pi i}\mathrm{Tr}\ln (P_{\pm }^{A}{P_{\pm }^{B}}^{\dag}),
\end{equation}%
where $U(T)\pm \mathbb{I}=U_{\pm }S^{\prime }_{\pm}V_{\pm }^{\dag }$, $%
P_{\pm }^{A}=U_{\pm }^{\dag }PU_{\pm }$, $P_{\pm }^{B}=V_{\pm }^{\dag
}PV_{\pm }$, and $P$ is defined in Eq.\eqref{P}. As depicted in Figs.~\ref{tv44}%
(c,d), the zero-mode singular value can be well characterized by the topological invariants. The number of zero-mode singular values equals $2\mathcal{V}_{-}$ (for $0$-modes) and $2\mathcal{V}_{+}$ (for $\pi $-modes), providing a direct count of
topologically protected corner states. By analyzing the singular value spectrum of simple Hermitian operators $\tilde{H}_{\pm }^{\prime}$ derived from $U(T)$, we obtain a new method to classify Floquet topological
phases.

\section{Discussion and Conclusion }
In summary, we investigated the higher-order topology of static and Floquet non-Hermitian SSH model.  With the help of singular value spectrum , we demonstrate that the zero-mode singular value robustly signals the emergence of topologically protected corner states of  energy spectrum in the thermodynamic limit, even in the absence of chiral symmetry. The winding numbers in real space can be defined to count the number of stable zero-mode singular states. We have established a unified framework to restore the bulk-boundary correspondence in non-Hermitian higher-order topological phases, both in static and Floquet systems.

In recent years, non-Hermitian Hamiltonian has been realized in photonic quantum walks \cite{PhysRevLett.133.070801}, Photonic Quasicrystal \cite{PhysRevLett.132.263801}, cold Rydberg quantum gases \cite{zhang2025observationnonhermitiantopologycold}, and waveguides \cite{bk7q-6r9d}. Floquet higher-order topological phases has been observed in acoustic lattice \cite{shiyan1} and Rydberg atoms \cite{shiyan2}. Based on these
developments, we believe that our approach is experimentally
feasible.

 \section{Acknowledgments}This work is supported by National Natural Science Foundation (Grants No. 12405007 and NO. 12305011), Funds for Young Scientists of Chongqing Municipal Education Commission(Grant No. KJQN202400603 and N0. KJQN202500619), Natural Science Foundation of Chongqing (Grant No. CSTB2025NSCQ-GPX1265, No. CSTB2022NSCQ-MSX0316, CSTB2024NSCQ-MSX0736, and CSTB2025NSCQ-GPX1315), and Chongqing Natural Science Foundation Project (Grant No. CSTB2025NSCQ-LZX0142)

\bibliography{references}
\end{document}